\documentclass[12pt]{article}
\usepackage{amsmath,amssymb,cite}
\usepackage{graphicx}

\newcommand{\One}{{\hbox{{\rm 1{\hbox to 1.5pt{\hss\rm1}}}}}}

\newcommand\be{\begin{equation}}
\newcommand\ee{\end{equation}}
\newcommand\bea{\begin{eqnarray}}
\newcommand\eea{\end{eqnarray}}

\newcommand\ba{\(\begin{array}}
\newcommand\ea{\end{array}\)}

 \def\d{\delta}

 \def\a{\alpha}
 \def\b{\beta}
 
 \def\d{\delta}
 \def\e{\epsilon}

 \def\k{\kappa}

 \def\s{\sigma}
 \def\t{\tau}
 
 \def\z{\zeta }

 \def\o{\omega }
 
 \newcommand{\tr}{\text{tr }}
 \def\la{\left\langle}
 \def\ra{\right\rangle}
 \def\hf{\dfrac{1}{2}}

\makeatletter
\@addtoreset{equation}{section}
\makeatother
\newcommand{\resection}[1]{\setcounter{equation}{0}\section{#1}}

\newcommand{\cref}[1]{(\ref{#1})}

\allowdisplaybreaks
\begin{document}

\setlength{\oddsidemargin}{0cm}
\setlength{\baselineskip}{7mm}

\begin{titlepage}
\renewcommand{\thefootnote}{\fnsymbol{footnote}}

\vspace*{0cm}
    \begin{Large}
       \begin{center}
         {Boundary operators in the one-matrix model}
       \end{center}
    \end{Large}
\vspace{0.7cm}

\begin{center}
Jean-Emile B{\sc ourgine$^{1}$}\footnote
            {
e-mail address : 
jebourgine@sogang.ac.kr},
Goro I{\sc shiki$^{1}$}\footnote
            {
e-mail address : 
ishiki@post.kek.jp}
    {\sc and}
Chaiho R{\sc im$^{2}$}\footnote
           {
e-mail address : 
rimpine@sogang.ac.kr}\\
      
\vspace{0.7cm}                    
{\it Department of Physics$^{2}$ and 
  Center for Quantum Spacetime (CQUeST)$^{1,2}$
}\\
{\it Sogang University, Seoul 121-742, Korea}
\end{center}

\vspace{0.7cm}

\begin{abstract}
\noindent

The one matrix model is known to reproduce in the continuum limit the $(2,2p+1)$ minimal Liouville gravity. Recently, two of the authors have shown how to construct arbitrary critical boundary conditions within this matrix model. So far, between two such boundary conditions only one boundary operator was constructed. In this paper, we explain how to construct all the set of boundary operators that can be inserted. As a consistency check, we reproduce the corresponding Liouville boundary 2pt function from the matrix model correlator. In addition, we remark a connection between a matrix model relation and the boundary ground ring operator insertion in the continuum theory.

\end{abstract}
\vfill
\end{titlepage}
\vfil\eject

\setcounter{footnote}{0}


\section{Introduction}
The matrix models generate the ensemble of discretized two dimensional surfaces as a Feynman diagrammatic expansion, thus providing a discrete formulation of the Liouville gravity \cite{BK,DS,GM,D,K,DFK,Polyakov,KPZ,David,DK} (see also the reviews \cite{GM93,DFGZ}). In the case of the one (hermitian) matrix model (OMM), the matrix potential can be tuned in the continuum limit in order to achieve the $(2,2p+1)$ minimal Liouville gravity (MLG$(2,2p+1)$), which consists of matter, Liouville and ghosts fields. Deformations of the critical potential can also be introduced, which produce the KdV renormalization group flows between the critical points. In the Liouville gravity context, this picture corresponds to perturbing the action with vertex operators. As suggested in \cite{MSS}, the exact relation between the matrix model (KdV) and the Liouville gravity couplings involves a non-trivial resonance transformation due to the contact terms. This transformation was obtained at the first order in \cite{MSS}, and the identification of the bulk one and two point correlation functions was performed. In \cite{BZ}, A. Belavin and A. Zamolodchikov obtained the resonance transformation reproducing the three-point correlation function of the
MLG$(2,2p+1)$ from the matrix model. Their result leads to a conjecture for the explicit form of the resonance transformation to all orders \cite{BZ}. This conjecture was checked up to the fifth order \cite{BZ,GT}. It was also shown in \cite{BR} that this transformation also works for the bulk one-point correlation function on the disc.

To complete our understanding of the discrete formulation of the Liouville gravity, we have to study the realization of boundary conditions (BCs) and boundary operators. Such approach has already been taken in various matrix models, including the RSOS and $O(n)$ models \cite{b-on-sos, b-on-sos2,bloop} and the two matrix model \cite{Martinec:1991ht, b-two-matrix}. Recently, two of the authors constructed new boundaries for the OMM using additional vector fields, providing the general MLG$(2,2p+1)$ BCs in the continuum limit \cite{Ishiki:2010wb}. Such BCs, referred as FZZT brane \cite{FZZ,Teschner2000}, depend on two parameters and will be labeled $(s;\ell )$. The parameter $s$ is related to the boundary cosmological constant $\mu_B$ which determines the BC for the Liouville field, 
\begin{align} 
\mu_B(s)=\sqrt{{\mu}/{\sin(\pi b^2)}} \cosh (\pi b s) .
\label{muB}
\end{align}
The matter BC is given by the Cardy state $(1,\ell )$  where $\ell$ runs from one to $p$ \cite{CL} and we use the Kac notation. The matrix model construction of \cite{Ishiki:2010wb} was interpreted in \cite{Bourgine:2010ja} as the realization of a linear relation among FZZT branes. These results were also generalized to the two matrix model which provides the general MLG$(p,q)$.

Between the two BC $(s;\ell)$ and $(t;m)$, we have to introduce one of the following boundary operators of dimension one,
\begin{align} 
{}^{(s;\ell )}B_{1 k}^{(t;m)} 
= {}^{(s;\ell )}\left[e^{\beta_{1 k}\, \phi}  \Phi_{1 k} \right]^{(t;m)},\quad k=|\ell-m|+1,\cdots,\ell+m-1;2
\label{bcc operator}
\end{align}
where $\phi$ denotes the Liouville field and $\Phi_{1 k}$ the $(1,k)$ primary matter field and $k$ has increment of two. The Liouville charge $\beta_{1 k}$ is related to the dimension of $\Phi_{1k}$ through the KPZ relations \cite{KPZ,David,DK} and satisfies the Seiberg's bound ($k\leq p$),
\begin{equation}
\beta_{1 k} = (1+k)\dfrac{b}{2},\quad b^2=\dfrac{2}{2p+1} .
\end{equation}
The diffeomorphism invariance of the correlation functions imposes to integrate the boundary operators \cref{bcc operator} over the boundary. The three conformal Killing vectors on the disc allow to fix the position of three boundary operators with the appropriate ghost dressing. Since there are no remaining coordinate dependence, these correlation functions are referred as `correlation numbers'. So far from the matrix model, only correlators involving the ${}^{(s;\ell )}B_{1 \ell+m-1}^{(t;m)}$ operators were explicitly known. The purpose of this article is to construct the matrix model operators providing in the continuum limit the general boundary operators \cref{bcc operator}. This will be achieved by the introduction of powers of the matrix inside the correlators.

This paper is organized as follows. In the section \ref{One matrix model with vectors}, we briefly review the construction of the OMM boundaries using the vector description. Then, we discuss in section \ref{Boundary perturbation} the perturbation of the boundary term, and the general form of the matrix boundary operators. These operators will be explicitly determined in the section \ref{Boundary two-point correlation numbers} using the requirement of vanishing boundary two-point functions for two different operators in the MLG$(2,2p+1)$. As a consistency check, we recover the expression of the two-point functions for two identical operators. Finally, the section \ref{summary} is devoted to summary and discussions.

\section{Boundary conditions in the one matrix model}
\label{One matrix model with vectors}
We consider the OMM coupled to a vector model, with partition function
\begin{align}
e^Z=\int DM \prod_a Dv^{(a)\dagger} \prod_a  Dv^{(a)} \exp
\left( -\frac{N}{g}{\rm tr}V(M) 
        -\sum_{a,b}v^{(a)\dagger}\cdot \Xi^{(a,b)}(M) \cdot v^{(b)} \right),
\label{starting point}
\end{align}
where $M$ is the $N\times N$ hermitian matrix, and $V(M)$, $\Xi^{(a,b)}(M)$ are some polynomials of $M$. The vectors $v^{(a)}$ (and their hermitian conjugates $v^{(a)\dagger}$) belong to the fundamental (conjugate) representation of $U(N)$, and ``$\cdot$'' represents the contraction of the $N$ dimensional indices. The flavors indices $a$ and $b$ label the boundary condition, $a=(s;\ell)$. The coupling $g$ is related to the bulk cosmological constant $\kappa^2=1/g$ which controls the area fluctuation of the worldsheet.

The vectors allow to emphasize the role played by the boundary, but can be integrated out, and the partition function \cref{starting point} is reduced to the following expansion,
\begin{align}
Z=\sum_{h=0}^{\infty}Z_h, \quad e^{Z_0}=\int DM e^{-\frac{N}{g}{\rm tr}V(M)},\quad Z_{h\geq 1}=\frac{1}{h!} \left\langle \{-{\rm Tr}\log \Xi (M)\}^h \right\rangle_c.
\label{partition function with h holes}
\end{align}
Here, $\langle \cdots \rangle_c$ represents the connected part of the expectation value with respect to the matrix integration with potential $V(M)$ and ${\rm Tr}$ stands for the trace over both flavor and matrix indices. The correlators $Z_h$ describe a worldsheet with $h$ disconnected boundaries, or holes. They can be expanded in powers of $N$, $Z_h=\sum_{g=0}^{\infty}Z_h^g N^{\chi}$; each $Z_h^g$ describes a worldsheet with $g$ handles and $h$ holes, and $\chi=2-2g-h$ is the corresponding Euler characteristic. In this paper, we focus on the disk partition function ($h=1$) in the planar limit ($g=0$),
\begin{align}
Z_{\rm disk}=- \lim_{N\to\infty} \dfrac{1}{N}\langle {\rm Tr}\log \Xi(M) \rangle .
\label{disk partition function}
\end{align}

In the Feynman diagrammatic expansion of the integral (\ref{starting point}), boundaries are formed by loops of the vectors, each flavor being associated to a different boundary \cite{Ishiki:2010wb}. The diagonal elements of the matrix $\Xi^{(a,b)}(M)$ create the corresponding boundaries, whereas the off-diagonal elements produce a flavor mixing leading to a boundary changing effect. When the matrix $\Xi^{(a,b)}(M)$ is purely diagonal with respect to the flavor indices, the correlator \cref{disk partition function} reduces to a sum over the disk correlators with different boundaries $a$, $Z_{\rm disk}=-\lim_{N\to\infty} \dfrac1N\sum_a \left\langle \tr\log{\Xi}^{(a,a)} \right\rangle$, where $\tr$ denotes the trace over the matrix indices.

The simplest, and most studied, case is to consider only a single boundary and define $\Xi (M)=M-x$. Its derivative with respect to the boundary cosmological constant $x$ is the well-known resolvent,
\begin{equation}
W(x)\equiv\lim_{N\to\infty}\dfrac{1}{N}\la\tr\dfrac1{x-M}\ra\equiv\hf V'(x)+\o(x).
\end{equation}
In order to achieve the $p$th critical point corresponding to the MLG$(2,2p+1)$, the potential $V(M)$ is fine-tuned in the continuum limit. The bulk cosmological constant gets renormalized, and we define $\epsilon^2\mu=\k-\k^*$ where $\k^*$ is the critical value for which the area blows up and the cut-off $\e$ is the lattice spacing. In this limit, the worldsheets with large boundaries dominate, and $x$ is sent to its critical value $x^*$. This value will be taken to be zero by a suitable shift of the matrix $M$, so that we define the renormalized boundary cosmological constant as $\e\xi=x$, which is identified with the parameter $\mu_B$ of the MLG$(2,2p+1)$. In this limit, the singular part of the resolvent $\o(x)$ is rescaled as $\omega(x)=\epsilon^{1/b^2}\omega(\xi)$, and has a branch over the support of the eigenvalue density $]-\infty,-u_0]$. This branch cut is resolved using the parameterization \cref{muB}, 
\begin{align}
\xi (s) = u_0 \cosh(\pi b s), \quad \omega(\xi)= u_0^{1/b^2} \cosh\left( \frac{\pi s}{b}  \right), \quad u_0=\sqrt{\mu /\sin(\pi b^2)}.
\label{resolvent}
\end{align}
The resolvent $\o(\xi)$ is identified with the disc boundary one point function of the dressed identity operator, with BC $(s;1)$ of the MLG$(2,2p+1)$.

The new matrix boundaries introduced in \cite{Ishiki:2010wb} are defined as $\Xi(M)=F_{\ell }(x,M)$ with 
\begin{align}
F_{\ell }(x,M)\equiv \prod_{j=-(\ell -1),2}^{\ell -1}(M-x_j),
\label{critical value of F}
\end{align}
where we denoted $x(s)=\e u_0\cosh{(\pi bs)}$ and 
\begin{align}
x_j(s)=\epsilon u_0\cosh (\pi b s+i \pi b^2 j ),\quad j=-(\ell-1),\cdots,(\ell-1);2.
\label{correct solution}
\end{align}
In the continuum limit, these couplings get renormalized as $\e\xi=x$, $\e\xi_j=x_j$, and the critical resolvant $\o$ takes the same value over the variables $\xi_j$, $\o(\xi_j)=(-1)^{\ell-1}\o(\xi)$. In this limit, the matrix operator $F_{\ell}(x,M)$ creates a boundary with $(s;\ell )$ BC, i.e. with a matter BC given by the Cardy state $(1,\ell)$ and a renormalized boundary cosmological constant $\xi(s)$. This interpretation follows from the study of the boundary one and two point correlators \cite{Ishiki:2010wb}, and more generally from a linear relation among FZZT branes \cite{Bourgine:2010ja}.

\section{Boundary perturbations}
\label{Boundary perturbation}
In this section, we study the perturbation away from the critical boundary constructed by $F_\ell(x,M)$. The bulk perturbations are introduced as a modification of the potential $V(M)$,
\begin{align}
\frac{N}{g}
V(M)\equiv \s_p(M)+\sum_{j=0}^{p-1}t_j\sigma_j(M),
\label{bulk perturbation}
\end{align}
where the $\s_j$ are the $j$th critical potentials leading to the MLG$(2,2j+1)$ \cite{GM}. The couplings $t_j$ generate the KdV flows between the critical points of the matrix model; they are related to the perturbation of the MLG$(2,2p+1)$ with bulk vertex operators by the resonance transformation. In a similar way, we would like to write the boundary term in \cref{starting point} as a perturbation of a critical background $\Xi_*^{(a,b)}$,
\begin{align}
\Xi^{(a,b)} (M) \equiv \Xi_*^{(a,b)}(M) + 
\sum_{j}c^{(a,b)}_j P^{(a ,b)}_j(x^{(a)},x^{(b)},M),\quad \Xi_*^{(a,b)}(M)=\d_{a,b}F_{\ell_a}(x^{(a)},M).
\label{boundary perturbation}
\end{align}

Let us first consider the case of a single boundary $\Xi(M)$; the perturbed disc partition function corresponds to
\begin{align}
Z_{\rm disk}=-\lim_{N\to\infty}\dfrac1N\la
{\rm tr}\log \left(\Xi_*(M)+\sum_jc_jP_j(x,M)\right)\ra,
\label{1f}
\end{align}
where the $c_j$ are the perturbative couplings related to the operators $P_j(x,M)$ and $\Xi_*(M)=F_\ell(x,M)$. This correlator can be expanded in the perturbation series,
\begin{align}
\begin{split}
\la{\rm tr}\log \left(F_\ell(x,M)+\sum_jc_jP_j(x,M)\right)\ra=&-\langle {\rm tr}\log F_\ell(x,M)\rangle-\sum_jc_j\left\langle {\rm tr}\frac{P_j(x,M)}{F_\ell(x,M)}\right\rangle\\
+&\hf \sum_{j,k}{c_jc_k}\left\langle {\rm tr} 
\frac{P_j(x,M)P_k(x,M)}{F_\ell(x,M)^2}\right\rangle +\cdots .
\end{split}
\end{align}
and we recover in the planar limit a disc correlator with insertion of zero, one, two, etc boundary operators. These operators $P_j(x,M)$ should satisfy two requirements: they have to be local and scaling operators. The first requirement imposes that $P_j(x,M)$ is a polynomial in $M$. Indeed, the insertion of powers of the matrix corresponds to the insertion of a finite edge on the boundary, which is renormalized toward a point in the continuum limit. To understand the implication of the second requirement, let us recall that in the continuum limit, the critical matrix correlators obey a scaling property which allows to define their scaling dimension.\footnote{More precisely, let $D(t_i)$ be a matrix correlator depending on the couplings $t_i$. This correlator can be decomposed into a regular $D^{(r)}(t_i)$ and critical $d(t_i)$ part. In the continuum limit, the regular part, which is polynomial in at least one of the boundary cosmological constant, is non-universal and shall be discarded. If the couplings take the critical values $t_i^*$, we define the renormalized couplings as $\e^{\a_i}\t_i=t_i-t_i^*$ and correlator $\e^\a d(\t_i)=d(t_i)$ where $\a$ and $\a_i$ are the corresponding scaling dimensions. The scaling property writes $d(\t_i)=\rho^{-\a}d(\rho^{\a_i}\t_i),\ \forall\rho$.} A matrix operator has scaling dimension $j$ if its introduction within matrix correlators modifies their scaling dimension from $\a$ to $\a+j$. Because of the shift of the matrix that imposed $x^*=0$, the matrix powers $M^j$ are scaling operators of dimension $j$ (see appendix \ref{Scaling property}). In the continuum limit, there exist only two scale parameters, given by the bulk and boundary cosmological constants. Thus, imposing that $P_j(x,M)$ is a monic polynomial in $M$ with scaling dimension $j$, it is restricted to be of the form
\begin{align}
P_j(x,M)=M^j+\sum_{n=1}^{j}\sum_{m=0}^{[n/2]} 
h_{nm}x^{n-2m}(\kappa -\kappa^*)^{m}M^{j-n},
\label{pol}
\end{align}
where $h_{nm}$ are $c$-numbers and we discarded terms that do not contribute to the continuum limit. A natural choice is to take $P_j(x,M)=F_j(x,M)$ since this generates a boundary flow similar to the KdV flow introduced by the perturbation \cref{bulk perturbation}. When one of the coupling $c_j$ goes from zero toward infinity, the continuum boundary flows from $(s;\ell)$ to $(s;j)$.

We now consider the perturbation of the off-diagonal elements of $\Xi^{(a,b)}$ between the two boundaries $a=(s;\ell)$ and $b=(t;m)$. To simplify the notations, we denote the polynomials by $P^{(\ell ,m)}_j(x,y,M)$; they should be local, monic and of degree $j$. Inserted between two different boundaries, the boundary operators can now depend on three scaling parameters in the continuum limit, given by the bulk and the two boundary cosmological constants. However, in this case, there is no clear picture of boundary flows available at this stage. In the next section we will follow a practical approach and determine the boundary operators directly in the MLG$(2,2p+1)$ frame, i.e. after the resonance transformation. In this frame, the critical part of the matrix model correlators for a disc with two boundaries should satisfy the condition
\begin{align}
\left\langle
{\rm tr}\frac{1}{F_\ell (x,M)F_m(y,M) }
P^{(\ell m)}_j (x,y,M)P^{(m \ell)}_k(y,x,M)
\right\rangle=\d_{jk} O_{\ell m}^j(x,y)
\label{orthogonality}
\end{align}
where $O_{\ell m}^j(x,y)$ is proportional to the Liouville boundary two-point function in the continuum limit.\footnote{In the case of the Liouville gravity boundary two-point function, Liouville, matter and ghost contributions factorize. Only the Liouville part depends on the cosmological constants.} We will obtain the polynomials $P^{(\ell m)}_j (x,y,M)$ using the condition \cref{orthogonality} only when $j\neq k$. As a consistency check, we compute the correlators for $j=k$, and recover the Liouville boundary two-point function, the analysis of which shows that $c_j^{(a,b)}$ couple to the $^{(s;\ell)}B_{1,\ell+m-1-2j}^{(t;m)}$ boundary operator.

\section{Boundary two-point correlation numbers}
\label{Boundary two-point correlation numbers}
We now turn to the determination of the polynomials $P^{(\ell m)}_j (x,y,M)$. In the rest of this paper, we systematically drop out all the non-universal part of the matrix correlators since they are irrelevant in the continuum limit. In order to study the insertion of the matrix powers on the boundary created by $F_\ell(x,M)$ and $F_m(y,M)$, we introduce the ratios $g_j^{(\ell m)}(x,y)$,
\begin{align}\label{def_gjlm}
&\left\langle {\rm tr}\frac{M^j}{F_{\ell }(x,M)F_m(y,M) }\right\rangle
\equiv 
g_j^{(\ell m)}(x,y)O_{\ell m}^0(x,y).
\end{align}
The correlators $O_{\ell m}^0(x,y)$ were studied in \cite{Ishiki:2010wb}; they have scaling dimension $\a_{\ell m}^0=p+\frac12-(l+m-2)$ and can be expressed in the continuum limit in terms of the critical resolvent
\begin{align}
O^0_{\ell m}(\xi,\z)=\frac{[\ell +m-2]!}{[\ell -1]![m-1]!}
\frac{\omega(\xi)-\omega(\z_{l+m})}
{\prod_{j=-(\ell +m-2),2}^{\ell +m-2}(\xi_j-\z)}, 
\label{Olm}
\end{align}
where $\e^{\a_{\ell m}^0} O_{\ell m}^0(\xi,\z)=O_{\ell m}^0(x,y)$, $\e\xi=x$, $\e\z=y$ and we used the notations
\begin{align}
[n]!\equiv \prod_{k=1}^{n}[k],\quad [k]\equiv\dfrac{q^k-q^{-k}}{q-q^{-1}},\quad q\equiv e^{i\pi b^2}.
\end{align}
The expression \cref{Olm} is known to reproduce in the continuum limit the Liouville part of the $^{(s;\ell)}B_{1\ell+m-1}^{(t;m)}$ boundary two-point functions. The ratios $g_j^{(\ell m)}(x,y)$ can be computed recursively using the relation
\begin{align}\label{rec}
g_j^{(\ell m)}(x,y)O^0_{\ell m}(x,y)&=
\left\langle {\rm tr}
\frac{M^{j-1}(M -x_{\pm (\ell -1)}+x_{\pm (\ell -1)})}
{F_{\ell }(x,M)F_m(y,M) }
\right\rangle 
\nonumber\\
&=g_{j-1}^{(\ell -1m)}(x_{\mp 1},y) O^0_{\ell -1m}(x_{\mp 1},y)
+x_{\pm (\ell -1)}g_{j -1}^{(\ell m)}(x,y)O^0_{\ell m}(x,y).
\end{align}
From its definition \cref{def_gjlm}, $g_j^{(\ell m)}(x,y)$ is a scaling quantity of dimension $j$ and we can define the renormalized ratio $g_j^{(\ell m)}(x,y)=\epsilon^{j}g_j^{(\ell m)}(\xi, \zeta)$ in the continuum limit. Substituting the expression (\ref{Olm}) for $O_{\ell m}^0(\xi,\z)$, the previous recursion relation can be written as 
\begin{align}
g_j^{(\ell m)}(\xi,\z)=\xi_{\pm (\ell -1)}g_{j -1}^{(\ell m)}(\xi,\z)
-\frac{[\ell -1]}{[\ell +m-2]}
(\xi_{\pm (\ell +m-2)}-\z)g_{j-1}^{(\ell -1m)}(\xi_{\mp 1},\z).
\label{recursion}
\end{align}
We note that $g_0^{(\ell m)}(\xi,\z)=1$, and $g_j^{(\ell m)}(\xi,\z)$ depends on the renormalized cosmological constants of the boundary $\xi$ and $\z$, and of the bulk through $u_0^2\propto\mu$; it is a polynomial of $\xi,\z$ with degree $j$.\footnote{The ratios $g_j^{(\ell m)}$ are obviously polynomials in $\z$ of degree $j$ for all $\ell,m$. The dependence on $\xi$ follows from the symmetry $g_j^{(\ell m)}(\xi,\z)=g_j^{(m \ell)}(\z,\xi)$.} The first two ratios are explicitly given by
\begin{align}
g_1^{(\ell m)}(\xi ,\zeta)&=\frac{[m-1]\xi+[\ell -1]\zeta}
{[\ell +m-2]}, \nonumber\\
g_2^{(\ell m)}(\xi ,\zeta)&=\frac{[\ell -1][\ell -2]\xi^2+
[m-1][m-2]\zeta^2}{[\ell +m-2][\ell +m-3]},
\nonumber\\                  & \;
+\frac{[2][\ell -1][m-1]\xi \zeta }{[\ell +m-2][\ell +m-3]}
+u_0^2\sin^2 \pi b^2\frac{[\ell -1][m-1]}{[\ell +m-3]} .
\end{align}

Now that the correlators \cref{def_gjlm} are determined, we investigate the consequence of the orthogonality condition \cref{orthogonality} for $j\neq k$. We first write the polynomials $P_j^{(\ell m)}$ as a sum of monomials,
\begin{align}
P^{(\ell m)}_j(x,y,M)=\sum_{k=0}^{j}a_{jk}^{(\ell m)}(x,y)M^k
\label{expansion}
\end{align}
with $a_{jj}^{(\ell m)}=1$. The orthogonality condition can be recursively linearized, leading to
\begin{equation}\label{orthogonality 2}
\sum_{k=0}^j g_{k+i}^{(\ell m)}(x,y) a_{jk}^{(\ell m)}(x,y)=\d_{ij} d_j^{(\ell m)}(x,y), (i\leq j)
\end{equation}
where we used the symmetry $P_j^{(\ell m)}(x,y,M)=P_j^{(m \ell)}(y,x,M)$, and introduced the ratios $d_j^{(\ell m)}(x,y)$,
\begin{align}
&O^j_{\ell m}(x,y) 
\equiv d_j^{(\ell m)}(x,y)O^0_{\ell m}(x,y).
\label{definition of d}
\end{align}
The system of equations (\ref{orthogonality 2}) being linear, there is a unique solution for $a_{jk}^{(\ell m)}$ and $d_j^{(\ell m)}$ in terms of $g^{(\ell m)}_k$, which is given by
\begin{equation}\label{ajk}
a_{jk}^{(\ell m)}=(-1)^{j+k}\dfrac{A_j^{(\ell m)\setminus k}}{A_{j-1}^{(\ell m)}},\quad d_j^{(\ell m)}=\frac{A_j^{(\ell m)}}{A_{j-1}^{(\ell m)}}
\end{equation}
where $A_j^{(\ell m)}$ is the following determinant,
\begin{align}\label{d=A/A}
A_j^{(\ell m)}
={\rm det} \left(
\begin{array}{cccc}
g_0^{(\ell m)}  & g_1^{(\ell m)} & \cdots & g_j^{(\ell m)}       \\
g_1^{(\ell m)}  & g_2^{(\ell m)} & \cdots & g_{j+1}^{(\ell m)}   \\
\vdots      & \vdots     &        & \vdots           \\
g_j^{(\ell m)}  & g_{j+1}^{(\ell m)} & \cdots & g_{2j}^{(\ell m)}   \\
\end{array}
\right).
\end{align}
and $A_j^{(\ell m)\setminus k}$ its first minor with respect to the last row and the $(k+1)$th column. The sum over the monomials \cref{expansion} can be performed, leading to a compact expression for the polynomials
\begin{align}
P^{(\ell m)}_j(x,y,z)=
\frac{1}{A_{j-1}^{(\ell m)}}
{\rm det}
\left(
\begin{array}{cccc}
g_0^{(\ell m)}     & g_1^{(\ell m)}     & \cdots & g_{j}^{(\ell m)} \\
g_1^{(\ell m)}     & g_2^{(\ell m)}     & \cdots & g_{j+1}^{(\ell m)} \\
\vdots             & \vdots             &        & \vdots  \\
g_{j-1}^{(\ell m)} & g_{j}^{(\ell m)}   & \cdots & g_{2j-1}^{(\ell m)} \\
1 & z & \cdots & z^{j} \\
\end{array}
\right).
\label{expression for P}
\end{align}
It can be shown that the coefficients $a_{jk}^{(\ell m)}$ and $d_j^{(\ell m)}$ are scaling variables of dimension $j-k$ and $2j$, respectively. In the continuum limit, $a_{jk}^{(\ell m)}(\xi,\z)$ and $d_j^{(\ell m)}(\xi,\z)$ are polynomials in $\xi$ and $\z$ of degree $j-k$ and $2j$.\footnote{The determinant $A_j^{(\ell m)}$ is a polynomial in $\z$ of degree $j(j+1)$, using the recursion relation \cref{recursion}, we can show that it has zeros for $\z=\xi_{\pm(l+m-2k)}$, $k=1\cdots j$ with multiplicity $j+1-k$. The same procedure also applies for the numerators of \cref{ajk} which are polynomial of degree $j(j+2)-k$, leading to the same zeros.} The scaling operators $P^{(\ell m)}_j(x,y,z)$ get renormalized into $\epsilon^j P^{(\ell m)}_j(\xi,\zeta ,\eta)$, where $\epsilon \xi=x$, $\epsilon \zeta=y$ and $\epsilon \eta=z$, and the first polynomials are given by
\begin{align}
P^{(\ell m)}_0(\xi,\zeta,\eta)&=1, \;\;\;P^{(\ell m)}_1(\xi ,\zeta ,\eta)=
\eta-\frac{[m-1]\xi+[\ell -1]\zeta}{[\ell +m-2]}, \nonumber\\
P^{(\ell m)}_2(\xi,\zeta,\eta)&=\eta^2
-\frac{[2]([m-2]\xi+[\ell -2]\zeta)}{[\ell +m-4]}\eta
-u_0^2 \sin^2 \pi b^2 \frac{[\ell -1][m-1]}{[\ell+m-3]}
\nonumber\\
&+\frac{[m-1][m-2]\xi^2+[\ell -1][\ell -2]\zeta^2+[2][\ell -2][m-2]\xi\zeta}
{[\ell +m-3][\ell+m-4]}
\end{align}

We now show that the solution we have found for $O_{\ell m}^j$ with the polynomials $P^{(\ell m)}_j$ previously determined indeed reproduce the Liouville part of the boundary two-point function. Unfortunately, the expression given for $d_j^{(\ell m)}$ is not very handy, so that we will take a different approach and work directly in the continuum limit where $d_j^{(\ell m)}(\xi,\z)$ is polynomial. We first notice that applying the recursion relation (\ref{recursion}) to (\ref{orthogonality 2}) we obtain the identity
\begin{align}
\sum_{k=0}^j g_{k+i-n}^{(\ell -nm)}(\xi_{\mp n},\z) 
a_{jk}^{(\ell m)}(\xi,\z)=0, \;\;\; (i=n, n+1 ,\cdots, j-1),
\label{orthogonality 3}
\end{align}
where $n$ is a non-negative integer less than $j$. Similarly, if we apply (\ref{recursion}) to (\ref{orthogonality 2}) with $i=j$ and eliminate the term proportional 
to $g^{(\ell m)}$ using the previous relation with $n=0$, we get
\begin{align}
d_j^{(\ell m)}(\xi,\z)&= -\frac{[\ell -1]}{[\ell +m-2]}
(\xi_{\pm (\ell +m-2)}-\z) \sum_{k=0}^j g_{k+j-1}^{(\ell -1m)}(\xi_{\mp 1},\z) 
a_{jk}^{(\ell m)}(\xi,\z).
\end{align}
Repeating this process we find that $d_j^{(\ell m)}$ has zeros at $\z=\xi_{\pm(\ell +m-2k)}$ for $k=1,2,\cdots,j$. Since its degree is $2j$, its functional form in $\z$ is completely determined as
\begin{align}
d_j^{(\ell m)}(\xi,\z)= 
C_j^{(\ell m)} 
\prod_{k=1}^j (\xi_{\ell +m-2k}-\z)(\xi_{-(\ell +m-2k)}-\z).
\end{align}
The remaining coefficient is independent of $\xi$ and $\z$; it is obtained in the appendix \ref{ap asymptotic form} from the asymptotic form of the recursion relation \cref{recursion} when $\z$ goes to infinity. This determines $d_j^{(\ell m)}$, and we finally find the the expression of $O^j_{\ell m}$,
\begin{align}
O^j_{\ell m}(\xi,\z)=(-1)^j\frac{[j]![\ell +m-2j-1]!}{[\ell +m-j-1]!}
O^0_{\ell -jm-j}(\xi,\z).
\label{final answer}
\end{align}
By substituting the expressions of \cref{correct solution}, and of the resolvent (\ref{resolvent}) in the continuum limit, we verify that this quantity is proportional to the MLG$(2,2p+1)$ boundary two-point function $\langle {}^{(s,\ell )}B_{1\ell +m-1-2j}^{(t,m)}B_{1\ell +m-1-2j}^{(s,\ell)}\rangle$ (see appendix A of the first reference in \cite{bloop}).\footnote{The precise identification with the Liouville Gravity correlation function involves ghost and matter contributions that can be trivialy obtained from a cosmological constants independent renormalization of the polynomials $P_j^{(lm)}$.}

\section{Summary and discussion}
\label{summary}

In this paper, we derived the matrix realization of arbitrary boundary operators using a polynomial insertion inside the correlators. These polynomials were determined directly in the CFT frame by imposing the orthogonality condition for the two-point function \cref{orthogonality}. The insertion of the polynomial $P_j(x,y,M)$ of degree $j$ given in \cref{expression for P} leads to consider the boundary operator ${}^{(s,\ell )}B_{1\ell +m-1-2j}^{(t,m)}$ instead of ${}^{(s,\ell )}B_{1\ell +m-1}^{(t,m)}$  inserted  between $(s;\ell)$ and $(t,m)$ boundaries. As a consistency check, we recovered the expression for the Liouville boundary two-point function coupled to a minimal model. Note that the case of operators inserted between two identical boundaries is obtained in the (non-singular) limit of equal boundary labels, $(s;\ell)=(t;m)$. 

The recursion relation \cref{rec} is one of the main identities we employed to derive the expression of the matrix operators. In the continuum limit, this relation can be interpreted as the insertion of two boundary ground ring operators $^sA_-^{s\pm ib}A_-^s$, as shown in appendix C. It would be interesting to study the realization of the ring structure with the one matrix model in more details.

Finally, we emphasize that the expression we found for the boundary operators was obtained for the disc with two different boundaries. It should be noted that because of the presence of resonant terms, we may have to introduce corrections to this expression when the operators are inserted into correlators describing a higher genus or a higher number of boundaries (possibly disconnected). This is the main open question we hope to address in a near future.

\section*{Acknowledgements}
We are indebted to I. Kostov for valuable discussions. 
This work is partially supported 
by the National Research Foundation of Korea (KNRF) 
grant funded by the Korea government (MEST)
2005-0049409 (B, I \& R) and R01-2008-000-21026-0 (R). 

\appendix

\resection{Scaling property}
\label{Scaling property}
In this appendix, we show that the matrix powers $M^j$ are scaling operators of dimension $j$. Let us first consider the correlators
\begin{align}
D_n(x^{(1)},\cdots,x^{(n)})=\left\langle {\rm tr} \prod_{i=1}^{n} \frac{1}{F_{\ell_i}(x^{(i)},M)}
 \right\rangle.
\label{n correlator}
\end{align}
One can be shown that these correlators have the scaling dimension $p+\frac{3}{2}-\sum_{i=1}^{n}\ell_i$ using recursively the identity
\begin{align}
\frac{1}{(M-a)(M-b)}=\frac{1}{a-b}
\left( \frac{1}{M-a} -\frac{1}{M-b} \right).
\end{align}
and the dimension $p+\frac12$ for the resolvant at the $p$-th critical point. Then, we consider an insertion of $M$ in the trace in (\ref{n correlator}), and use the relation
\begin{align}
\frac{M}{F_{\ell}(x,M)}=\frac{1}{F_{\ell-1}(x_{\mp 1},M)}
+\frac{x_{\pm (\ell -1)}}{F_{\ell }(x,M)},
\end{align}
to decompose this correlator as
\begin{align}
\left\langle 
{\rm tr} \prod_{i=1}^{n} \frac{M}{F_{\ell_i}(x^{(i)},M)}
\right\rangle
=
\left\langle 
{\rm tr} \prod_{i=1}^{n-1} \frac{1}{F_{\ell_i}(x^{(i)},M)}
\frac{1}{F_{\ell_n-1}(x_{\mp 1}^{(n)},M)}
\right\rangle
+x_{\pm (\ell-1)}^{(n)} 
\left\langle 
{\rm tr} \prod_{i=1}^{n} \frac{1}{F_{\ell_i}(x^{(i)},M)}
\right\rangle.
\end{align}
On the right-hand side, both terms have the dimension $p+\frac{5}{2}-\sum_{i=1}^{n}\ell_i$, so that the operator $M$ has scaling dimension one. This conclusion 
can be easily generalized to the insertion of the higher power of $M$, the operator $M^j$ leading to correlators of dimensions $p+\frac{3}{2}+j-\sum_{i=1}^{n}\ell_i$, i.e. it has dimension $j$.

\resection{Asymptotic form of $ d_j^{(\ell m)} $}
\label{ap asymptotic form}
In this appendix, we calculate $d_j^{(\ell m)}(\xi,\z)$ in the large $\z$ limit. In this limit, the ratios $g_j^{(\ell m)}$ are independent of $\xi$ and the relation (\ref{recursion}) can be easily solved,
\begin{align}
g_j^{(\ell m)}(\z)=\z^j \prod_{k=1}^j \frac{[\ell -k]}{[\ell+m-1-k]}.
\label{g in large y}
\end{align}
By substituting (\ref{g in large y}),
$A_j^{(\ell m)}$ defined in (\ref{d=A/A}) is written as
\begin{align}
A_j^{(\ell m)}=\z^{j^2+j}\prod_{k=1}^j
\left( \frac{[\ell -k]}{[\ell +m-1-k]}\right)^{j+1-k}B_j^{(\ell m)},
\end{align}
where $B_j^{(\ell m)}$ is a determinant of a 
$(j+1)\times (j+1)$
matrix given by,
\begin{align}
B_j^{(\ell m)}={\rm det}
\left(
\begin{array}{ccccc}
1  & \frac{[\ell -1]}{[\ell +m-2]}
   & \prod_{k=1}^2\frac{[\ell -k]}{[\ell +m-1-k]}  
   & \cdots               
   & \prod_{k=1}^j\frac{[\ell -k]}{[\ell +m-1-k]} \\
1  & \frac{[\ell -2]}{[\ell +m-3]}
   & \prod_{k=1}^2\frac{[\ell -1-k]}{[\ell +m-2-k]}   
   & \cdots               
   & \prod_{k=1}^j\frac{[\ell -1-k]}{[\ell +m-2-k]} \\
\vdots      
   & \vdots   
   & \vdots 
   &    
   & \vdots           \\
1  & \frac{[\ell -1-j]}{[\ell +m-2-j]} 
   & \prod_{k=1}^2\frac{[\ell -j-k]}{[\ell +m-1-j-k]} 
   & \cdots               
   & \prod_{k=1}^j\frac{[\ell -j-k]}{[\ell +m-1-j-k]} \\
\end{array}
\right).
\label{def of B}
\end{align}
$B_j^{(\ell m)}$ satisfies the following recursion relation,
\begin{align}
B_j^{(\ell m)}=[m-1]^j(-1)^j \prod_{k=1}^j 
\frac{[k]}{[\ell +m-1-k][\ell +m-2-k]} 
B_{j-1}^{(\ell -1m-1)}.
\label{recursion for B}
\end{align}
This is obtained by subtracting $i+1$-th row 
from $i$-th row for $i=1,2,\cdots,j$
in the determinant in (\ref{def of B})
and using the identity,
\begin{align}
[a+c][b]-[a][b+c]=[b-a][c].
\end{align}
Solving (\ref{recursion for B}) with $B_0^{(\ell m)}=1$, 
we find
\begin{align}
A_j^{(\ell m)}&=(-1)^{\frac{j(j+1)}{2}}\z^{j(j+1)}\prod_{k=1}^j
\left(\frac{[\ell -k][m-k]}{[\ell +m-1-k]}\right)^{j+1-k}
\nonumber\\ 
&\times \prod_{q=1}^j \prod_{k=1}^{j+1-q}
\left( \frac{[k]}{[\ell +m+1-2q-k][\ell +m-2q-k]}  \right).
\end{align}
By taking the ratio as in (\ref{d=A/A}), 
we get the asymptotic form of $d_j^{(\ell m)}$:
\begin{align}
d_j^{(\ell m)}(\z)=(-1)^j \z^{2j} \prod_{k=1}^j 
\left( 
\frac{[k][\ell -k][m-k]}{[\ell +m-1-k][\ell +m-j-k][\ell +m-j-1-k]}
\right).
\end{align}

\resection{Boundary ground ring}
The ground ring structure discovered in \cite{Witten1992} was investigated within the minimal string theory in \cite{Govindarajan1992,Govindarajan1993}. It was then generalized to the boundary by I. Kostov in \cite{Kostov2004,Kostov2004a}, and we employ here the notations of \cite{Kostov2004}. On the boundary, the ring is generated by two operators denoted $A_\pm$ whose coordinate derivatives are BRST-exact, allowing to move them freely on the boundary. These operators are built over degenerate matter and Liouville operators and have truncated OPE with the vertex operators. Below, only the $A_-$ operator, build over the $\Phi_{1,2}$ matter primary field, will play a role. The operator $^{s_1}A_-^{s_2}$ can be inserted between two boundaries with parameters satisfying $s_1-s_2=\pm i b$ or $s_1+s_2=\pm ib$.

Two different dressed boundary vertex operators can be built, depending on the choice of the root of the KPZ equation, $B_P^{(\pm)}$ where $P$ denotes the Liouville momentum related to the Liouville charge by $\b=Q/2-|P|$. Physically realized boundary operators are known to be $B_P^{(+)}$ when $P>0$ and $B_P^{(-)}$ when $P<0$. We suppose here that the operators have a negative momentum, $P_{1,\ell}=\ell b/2-1/(2b)$, so that we only need to consider the two following fusion relations between the boundary ground ring (bgr) and vertex operators, which simply writes within a suitable normalization
\begin{equation}\label{rel_fusion}
^{s_1\pm ib}A_- ^{s_1}B_P^{(-)\ s_2}=^{s_1}B_{P-b/2}^{(-)\ s_2},\quad ^{s_1}B_P^{(-)\ s_2\pm ib}A_- ^{s_2}=-^{s_1}B_{P-b/2}^{(-)\ s_2}.
\end{equation}
Note that the free field ($\mu=\mu_B=0$) fusion relations are deformed by the Liouville potential, leading to complicated fusion relations in the presence of integrated boundary operators. Similarly, the matter screening charge arising from the Coulomb gas representation of the minimal model further modifies these relations. However, such modifications only arise when considering the fusion of $A_-$ with vertex operators $B_P^{(+)}$, and the relations \cref{rel_fusion} are exact.

\begin{figure}[t]
\centering
\includegraphics{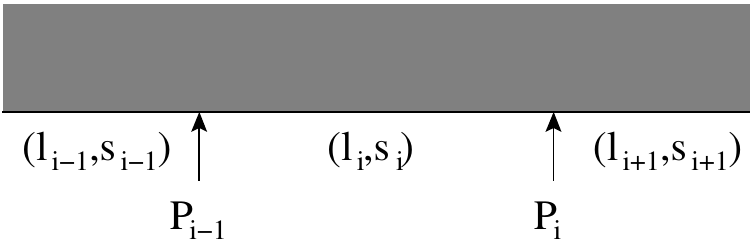}
\caption{Detail of the disc boundary.}
\label{fig1}
\end{figure}

\begin{figure}[t]
\centering
\includegraphics{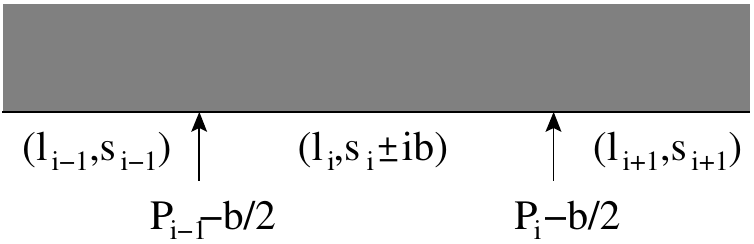}
\caption{Detail of the disc boundary modified by the matrix insertion.}
\label{fig2}
\end{figure}

\begin{figure}[t]
\centering
\includegraphics{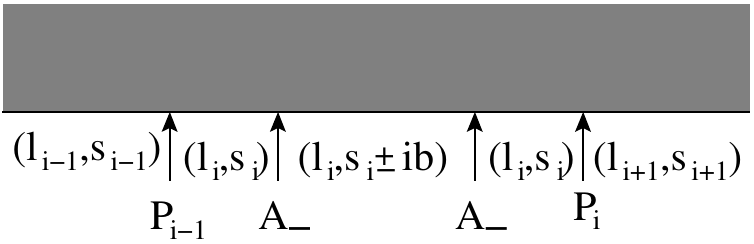}
\caption{Detail of the disc boundary modified by two bgr operator insertions.}
\label{fig3}
\end{figure}

In this appendix, we interpret the insertion of the monomial $(x_{\mp(\ell_i-1)}^{(i)}-M)$ on the boundary constructed by $F_{\ell_i}(x^{(i)})$ as two insertions of the bgr operator $A_-$ on the FZZT brane $(\ell_i,s_i)$. We consider the correlator $D_n$ defined in (\ref{n correlator}). In the continuum limit, the matrix operator $F_{\ell_i}(x^{(i)})$ creates the boundary $(\ell_i,s_i)$, and between boundaries $(\ell_i,s_i)$ and $(\ell_{i+1},s_{i+1})$, the boundary operator $^{s_i}B_{P_i}^{(-)\ s_{i+1}}$ should be inserted, with momentum $P_i=P_{1,\ell_i+\ell_{i+1}-1}$ (see figure \ref{fig1}). This identification does not involve any resonance term since the MLG coupling has maximal scaling so it appears only once in the resonance transformation.\footnote{On the matrix model side, all the operators $F_{\ell_i}$ are commuting, leading to conjecture that the order of the boundaries does not matter in the Liouville gravity side. More precisely,
\begin{align}
\begin{split}
&\la\tr\dfrac1{F_{\ell_1}}\cdots\dfrac1{F_{\ell_i}}\cdots\dfrac1{F_{\ell_j}}\cdots\dfrac1{F_{\ell_n}}\ra=\la\tr\dfrac1{F_{\ell_1}}\cdots\dfrac1{F_{\ell_j}}\cdots\dfrac1{F_{\ell_i}}\cdots\dfrac1{F_{\ell_n}}\ra\\
\Rightarrow\quad &\la^{s_1}B_{P_1}^{s_2}\cdots ^{s_{i-1}}B_{P_{i-1}}^{s_{i}}B_{P_i}^{s_{i+1}}\cdots^{s_{j-1}}B_{P_{j-1}}^{s_{j}}B_{P_j}^{s_{j+1}}\cdots ^{s_n}B_{P_n}^{s_1}\ra\\
=&\la^{s_1}B_{P_1}^{s_2}\cdots ^{s_{i-1}}B_{\tilde{P}_{i-1}}^{s_{j}}B_{\tilde{P}_i}^{s_{i+1}}\cdots^{s_{j-1}}B_{\tilde{P}_{j-1}}^{s_{i}}B_{\tilde{P}_j}^{s_{j+1}}\cdots ^{s_n}B_{P_n}^{s_1}\ra
\end{split}
\end{align}
with $\tilde{P}_{i-1}=P_{1,\ell_{i-1}+\ell_j-1}$, $\tilde{P}_i=P_{1,\ell_j+\ell_{i+1}-1}$, $\tilde{P}_{j-1}=P_{1,\ell_{j-1}+\ell_i-1}$ and $\tilde{P}_i=P_{1,\ell_i+\ell_{j+1}-1}$. At the level of the 3pt function, this can be shown using the reflection property.} Let us now examine the relation
\begin{align}
\begin{split}\label{mat_bgr}
&\la\tr\dfrac1{F_{\ell_1}(x^{(1)})}\cdots\dfrac1{F_{\ell_n}(x^{(n)})}(x_{\mp(\ell_i-1)}^{(i)}-M)\ra\\
=&-\la\tr\dfrac1{F_{\ell_1}(x^{(1)})}\cdots\dfrac1{F_{\ell_{i-1}}(x^{(i-1)})}\dfrac1{F_{\ell_i-1}(x_{\pm1}^{(i)})}\dfrac1{F_{\ell_{i+1}}(x^{(i+1)})}\cdots\dfrac1{F_{\ell_n}(x^{(n)})}\ra
\end{split}
\end{align}
From the RHS, we see that the effect of $(x_{\mp(\ell_i-1)}^{(i)}-M)$ is to modify the criticality of the $i$th boundary, $\ell_i\to \ell_i-1$, thus shifting the momentum of the neighboring boundary operators ($P_{i-1}\to P_{i-1}-b/2$, $P_i\to P_i-b/2$), and to shift its boundary cosmological constant $s_i\to s_i\pm ib$ (see figure \ref{fig2}). These effects are exactly those obtained by the insertion of two bgr operators $^{s_i}A_-^{s_i\pm ib}A_-^{s_i}$ (see figure \ref{fig3}). Indeed, using the relations \cref{rel_fusion} to fuse the bgr operator to the nearby vertex operators, we exactly recover the continuum limit of the identity \cref{mat_bgr}.

\end{document}